\documentclass[12pt,epsf,epsfig,psfig]{article}
\usepackage{graphicx}
\usepackage{epsfig}
\usepackage{cite}
\def\q{q \bar q}

\def\be{\begin{equation}}
\def\ee{\end{equation}}

\def\NP{{ Nucl.\ Phys.\ }}
\def\PL{{ Phys.\ Lett.\ }}
\def\PR{{ Phys.\ Rev.\ }}

\def\PRL{{ Phys.\ Rev.\ Lett.\ }}

\def\ZP{{ Z.\ Phys.\ }}
\def\EP{{ Eur.\ Phys.\ J.\ C}}

\begin{document}

~ \hfill BI-TP 2014/18

~~\vskip2cm

\centerline{\Large \bf Hadron Freeze-Out and Unruh Radiation}

\vskip1cm

\centerline{\bf Paolo Castorina$^{\rm a,b,}$, 
Alfredo Iorio$^{\rm c}$ and Helmut Satz$^{\rm d}$}

\bigskip
\centerline{a: Dipartimento di Fisica ed Astronomia, 
Universita' di Catania, Italy}
\centerline{b: INFN sezione di Catania, Catania, Italy}
\centerline{c: Faculty of Mathematics and Pysics, 
Charles Univeristy in Prague, Czech Republic}
\centerline{d: Fakult\"at f\"ur Physik, Universit\"at Bielefeld, Germany}

\vskip2cm

\centerline{\large \bf Abstract}

\bigskip

We consider hadron production in high energy collisions as an Unruh radiation 
phenomenon. This mechanism describes the production pattern of newly formed 
hadrons and is directly applicable at vanishing baryonchemical potential, 
$\mu \simeq 0$. It had already been found to correctly yield the hadronisation 
temperature, $T_h = \sqrt{\sigma/2 \pi} \simeq 165$ MeV in terms of the string
tension $\sigma$. Here we show that the Unruh mechanism also predicts hadronic 
freeze-out conditions, giving
 $s/T_h^3 = 3 \pi^2 /4 \simeq 7.4$ in terms of the entropy density $s$ and
$\langle E \rangle / \langle N \rangle = \sqrt{2 \pi \sigma} \simeq 1.09$ 
for the average energy per hadron. These predictions provide a theoretical 
basis for previous phenomenological results and are also in accord with
recent lattice studies.

\newpage

The relative abundances of hadrons produced in $e^+e^-$ annihilation
\cite{Beca-e,erice,Beca-h}, in hadron-hadron interactions \cite{Beca-p,Beca-h} 
and in the collisions of heavy nuclei 
\cite{HI-C,HI-CS,HI-R,HI-PBM,HI-BGS,HI-B1,HI-BRS}, 
are well described by an ideal hadronic resonance gas at fixed
temperature $T$ and baryonchemical potential $\mu$. In the high energy, 
low baryon density regime, from around 20 GeV up to the TeV range, such 
a description leads to a universal chemical freeze-out temperature $T_H 
\simeq 165 \pm 5$ MeV \cite{Beca-Biele}. This temperature moreover
agrees well with lattice QCD results at low or vanishing $\mu$ \cite{hot}.
Extensions to larger baryon density have led to various phenomenological 
proposals for freeze-out conditions \cite{CR1,CR2,CR3,PS,MS,Cley,Tawfik}. 
While these models indeed describe the present data quite well, neither 
the crucial parameters nor the conceptual forms appear to have a basic 
theoretical justification. The aim of this paper is to show that for low
or vanishing $\mu$, such a basis indeed exists.  

\medskip

The universality of the observed hadronisation temperature suggests a common 
origin for all high energy collisions, and it was recently proposed \cite{CKS} 
that high energy thermal hadron production is the QCD counterpart of 
Unruh radiation \cite{Un}, emitted at the event horizon 
due to color confinement. Let us briefly recall the relevant formalism.

\medskip

\begin{figure}[h]
\centerline{\psfig{file=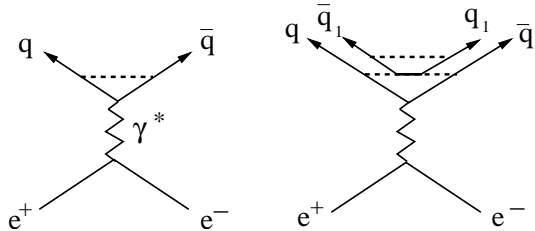,width=7cm} }
\caption{Quark formation in $e^+e^-$ annihilation}
\label{anni}
\end{figure}

The basic mechanism of Unruh radiation is tunnelling through the confining 
event horizon. It is most simply illustrated by hadron production through 
$e^+ e^-$ annihilation into a $\q$ pair; see Fig. \ref{anni}. 

The attempt to separate ends for the initial $\q$ pair at a distance $R$, 
when each quark hits the confinement horizon, i.e., when it reaches
the end of the binding string; the separation can now continue only 
if a further quark-antiquark system is excited from the vacuum. Although 
the new pair $q_1 \bar{q}_1$ is at rest in the overall center of mass system, 
each of its constituents has a transverse momentum $k_T$, determined by the 
uncertainty relation in terms of the transverse dimension of the string 
flux tube. String theory \cite{Lue} gives for the basic thickness
\be
r_T=\sqrt{2/\pi \sigma},
\label{1}
\ee
leading to
\be
k_T=\sqrt{\pi \sigma/2}.
\label{2}
\ee
The maximum separation distance $R$ is thus specified by
\be
\sigma R = 2 \sqrt{m_q^2 + k_T^2} = 2k_T,
\label{3}
\ee
where we have taken $m_q=0$ for the quark mass. From this we obtain
\be
R = \sqrt{2 \pi / \sigma}
\label{4}
\ee
as the string breaking distance. The departing quark $q$ now pulls the 
newly formed $\bar q_1$ along, giving it an 
acceleration 
\cite{CKS,BCMS}
\be
a = \sqrt{2 \pi \sigma}.
\label{5}
\ee 
The $q_1\bar q_1$ pair eventually suffers the same fate as the $\q$ pair: 
it is separated up to its confinement horizon, where it again excites a new
pair, which is now emitted as Unruh radiation of temperature
\be
T_h = {a / 2\pi} = \sqrt{\sigma/2 \pi}.
\label{6}
\ee
This process is sequentially repeated until the energy of the initial
``driving'' quarks $q$ and $\bar q$ is exhausted. 

\begin{figure}[h]
\centerline{\psfig{file=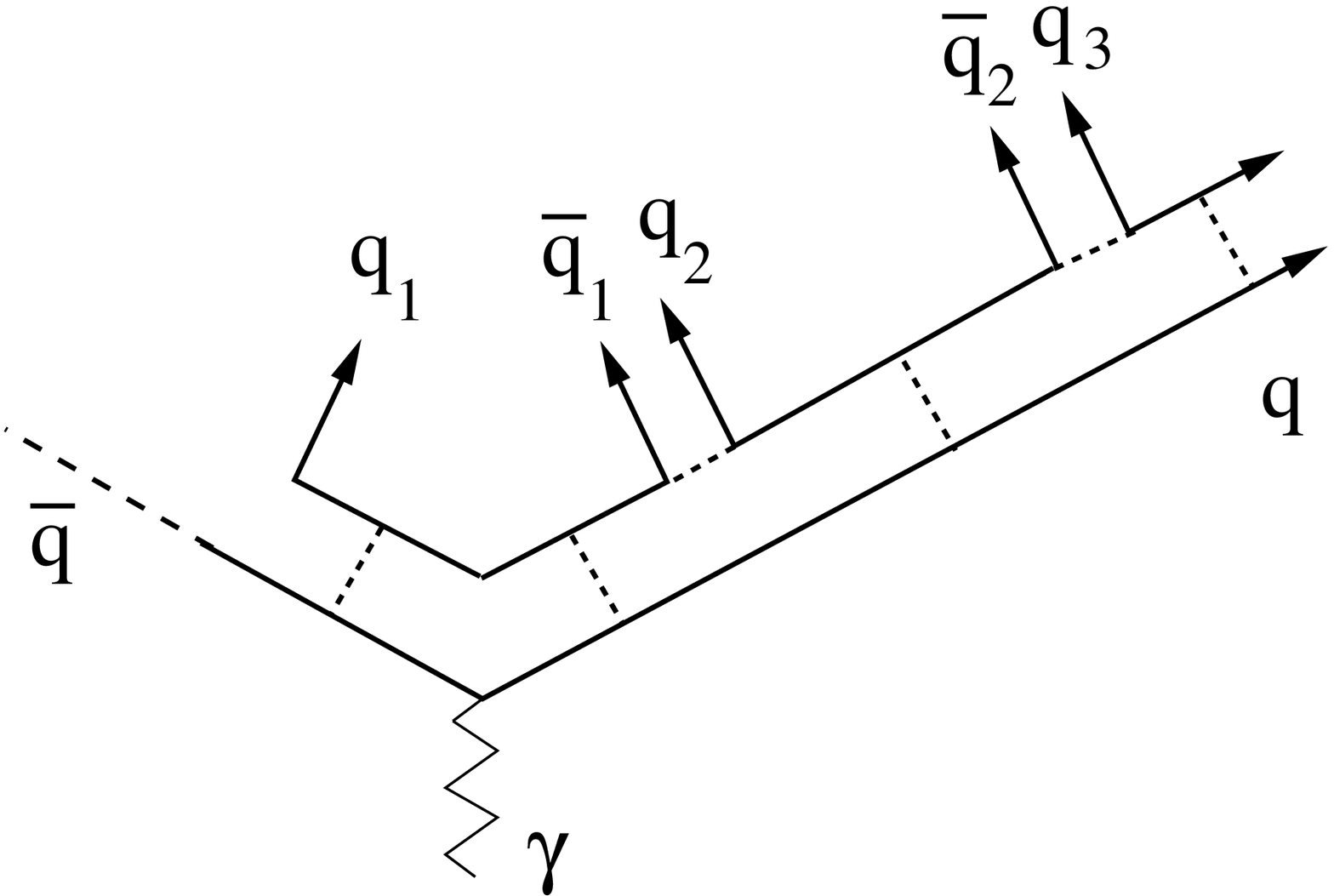,width=5cm}}
\caption{Sequential quark formation in $e^+e^-$ annihilation}
\label{anni2}
\end{figure}

\medskip

The case of $e^+e^-$ annihilation corresponds to baryochemical potential,
$\mu=0$. Here one finds the average value $\sigma \simeq 0.19 \pm 0.03$ 
GeV$^2$ \cite{tens1,tens2,tens3,tens4,tens5}, which with eq.\ (\ref{6})
then leads to 
\be
T_h(\mu=0)  = \sqrt{\sigma/2 \pi} \simeq 175 \pm 15~{\rm MeV}.
\label{7}
\ee
for the freeze-out temperature at $\mu=0$, in good agreement with the results
of both the mentioned resonance gas analyses \cite{CR1,CR2,CR3}
and lattice QCD studies \cite{hot}.
The fundamental mechanism in the Unruh scenario is quark acceleration leading 
to string breaking with the resuling pair production, as specified by eq.\ 
(\ref{3}). As long as we assume a vanishing quark mass, the only dimensional 
parameter in the entire formalism then is the string tension $\sigma$.

\medskip

The Unruh mechanism provides a theoretical basis for the production
of newly formed hadrons in high energy collisions. It does not address
the role of the nucleons already present in the initial state of heavy
ion collisions, and as such it can {\sl a priori} describe the 
freeze-out process only as long as there are no significant baryon-density 
effects, i.e., only for $\mu \simeq 0$\footnote{An attempt to address finite 
$\mu$ effects is given in \cite{cgi}.}
Here, however, it makes more 
well-defined predicitions than just eq.\ (\ref{7}).

\medskip

The energy of the pair produced by string breaking, i.e., of the newly
formed hadron, is from eq's.\ (\ref{2},\ref{3}) given by
\be
E_h = \sigma R = \sqrt{2 \pi \sigma}.
\label{8}
\ee
In the central rapidity region of high energy collisions, one has
$\mu \simeq 0$, so that $E_h$ is in fact the average energy 
$\langle E \rangle$ per hadron, with an average number $\langle N 
\rangle$ of newly produced hadrons.
Hence we obtain
\be
{\langle E \rangle \over \langle N \rangle
} = \sqrt{2 \pi \sigma} \simeq 1.09 \pm 0.08,
\label{9}  
\ee 
in accord with the phenomenological fit obtained from the species
abundances in high energy collisions \cite{CR1,CR2,CR3}.

\medskip

Next, we turn to the entropy. Because of the event horizon
caused by color confinement, this is necessarily an entanglement entropy,
caused by quantum field modes on both sides of the horizon. Its general
form is \cite{terashima,entaentropy}
\be 
\label{entropymatter}
S_{\rm ent} = \alpha \frac{A}{r^2} \;,
\ee
where $A$ is the area of the event horizon, $r$ the scale of 
characteristic quantum fluctuations and $\alpha$ an 
undetermined numerical constant. This expression shares the holographic 
structure\footnote{Holography of entanglement entropy is a quite general 
result, see \cite{srenidcky,solodukin}.} with the Bekenstein-Hawking 
entropy \cite{bekensteinEntropy,hawkingEntropy} for a black hole,
\be
S_{\rm BH} = {1 \over 4}~ {A_{\rm BH} \over r_P^2} ,
\label{10}
\ee
where $A_{\rm BH} = 4 \pi R_S^2$ denotes the surface area of the hole and
$R_S=2GM/c^2$ is its Schwarzschild radius. The quantity $r_P= \sqrt{\hbar
G/c^3}$ is the Planck length, setting the smallest possible fluctuation
scale. 

\medskip

\begin{figure}[h]
\centerline{\psfig{file=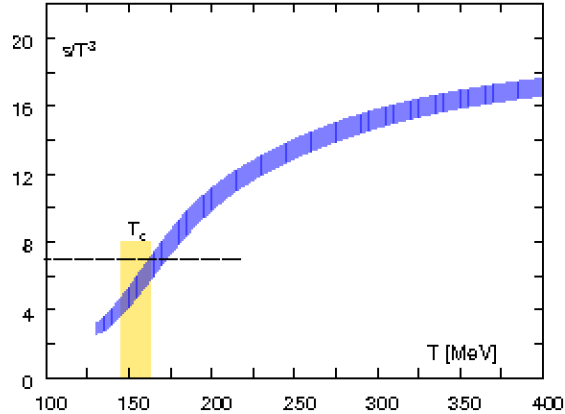,width=5.5cm,angle=-90}}
\caption{Lattice results \cite{hot} for $s/T^3$ as function of the temperature}
\label{swaga}
\end{figure}

The Bekenstein-Hawking relation (\ref{10}) also holds in the case of an
accelerated observer (Rindler spacetime), where it 
corresponds to the near-horizon approximation of a black hole 
spacetime \cite{laflamme}.
Here we take it to be valid also in our case, where gravity is not 
involved and the entire entropy is of the entanglement type. 
The scale of the characteristic quantum fluctuations is now given 
by eq.\ (\ref{1}), and we obtain
\be
S_h = {1\over 4}~ {A_h \over r_T^2} = {1\over 4}~ {4 \pi R^2 \over r_T^2}
\label{11}
\ee
for the entropy in hadron production. 
The parameter $R$ is given by eq.\ (\ref{4}), and the smallest 
fluctuation scale is the transverse string 
thickness (\ref{1}). Inserting these expressions into eq.\ (\ref{11}) 
gives for the entropy associated to hadron production
\be
S_h = \pi^3,
\label{S}
\ee
i.e., it becomes a pure number,
and also the entropy {\sl density} divided by $T^3$ remains dimensionles,
giving
\be
{s  \over  T^3} = {S_h \over (4 \pi/3) R^3 T^3} =   
{3 \pi^2 \over 4} \simeq 7.4
\label{entrop}
\ee
as freeze-out condition in terms of $s(T)$ and $T$. This result is in
accord with the value obtained for $s/T^3$ from species abundance analyses
in terms of the ideal resonance gas model \cite{Cley,Tawfik}. In Fig.\ 
\ref{swaga} we see that it agrees as well with the most recent
lattice QCD studies \cite{hot}.

\medskip

We close with two remarks. First, we comment on freeze-out for larger 
baryochemical potential $\mu$. This implies the presence of nucleons in 
the initial state, not
formed by the collision, and their interaction with the newly formed ones.
Such dynamical effects are not included in the Unruh formalism and hence
cannot be addressed here. We therefore cannot answer the tantalizing question
of why the conditions $\langle E / \rangle N \simeq 1.08$ or $s/T^3 = 
7.4$, valid for $\mu \simeq 0$, appear to remain as valid also for 
increasing $\mu$. This must imply an intricate interplay the decreasing role
of secondary mesons vs.\ the increasing role of the nucleons present already
in the initial state.    

\medskip

Secondly, we want to emphasize that besides the obvious importance of 
addressing issues of QCD phenomenology, the Unruh phenomenon appearing
here is of great importance in its own right. Unruh phenomena have eluded 
direct observation since decades, and having them here at our disposal in
the laboratory may well help in solving some of the long-standing open 
problems of theoretical aspects in fundamental physics, see, e.g., 
\cite{iorio}. One such issue might be the clarification of 
the entanglement nature of the Bekenstein-Hawking entropy.

\medskip

\centerline{\bf Acknowledgements}

\bigskip

We thank S.\ Mukherjee, BNL, for providing us with the lattice results of 
the HotQCD Collaboration (Fig.\ 3).
P.\ C.\ thanks the Faculty of Physics, Bielefeld University, for the kind
hospitality. 
A.\ I.\ wishes to thank the Department of Physics and Astronomy of Catania 
University, and INFN Sezione di Catania, for the kind hospitality, 
and acknowledges the Czech Science Foundation 
(GA\v{C}R), Contract No. 14-07983S, for support.
H.\ S.\ very belatedly thanks A. Tawfik for asking him many years ago
why hadron production data lead to $s/T^3=7$.

\bigskip


\end{document}